\documentclass[5p]{elsarticle}
\usepackage{lineno,hyperref}
\usepackage{mhchem}
\usepackage{mathtools}

\journal{Ultramicroscopy}
\usepackage[utf8]{inputenc}

\title{Simultaneous multi-region background subtraction for core-level EEL spectra}
\author[{1,cor}]{Jacob T. Held}
\author[1]{Hwanhui Yun}
\author[{1,cor}]{K. Andre Mkhoyan}

\address[1]{Department of Chemical Engineering and Materials Science, University of Minnesota, Minneapolis, Minnesota 55455, United States}
\cortext[cor]{Corresponding author: heldx123@umn.edu, mkhoyan@umn.edu}

\date{March 2019}

\begin{document}
\begin{abstract}
We present a multi-region extension of standard power-law background subtraction for core-level EEL spectra to improve the robustness of background removal. This method takes advantage of the post-edge shape of core-loss EEL edges to enable simultaneous and co-dependent fitting of pre- and post-edge background regions. This method also produces simultaneous and consistent background removal from multiple edges in a single EEL spectrum. The stability of this method with respect to the fitting energy window is also discussed.
\end{abstract}

\begin{keyword}
EELS \sep STEM \sep Core-Level \sep Background \sep Fitting
\end{keyword}
\maketitle

\newpage

\section{Introduction}
An important step in the preparation of core-level electron energy-loss (EEL) spectra for analysis and quantification in scanning transmission electron microscopy (STEM) is the removal of the background under the edge of interest. The most common practice to remove the background from core-loss edges is to fit a 10-30 eV pre-edge window with a single power law of the form $f(E)=A E^{-r}$, where $A$ is a scaling coefficient and $r$ defines the curvature of the background.\cite{egerton2011electron} The fit curve is then extrapolated and subtracted from the spectrum, leaving background-free core-loss features.

This approach generally works well for thin samples (minimizing bulk plasmon and multiple scattering contributions) with high signal-to-noise core-loss edges above ~100 eV (beyond the strong influence of the bulk plasmon) that do not overlap with other signals. However, in many cases, these conditions are not met, resulting in poor estimation of the background.\cite{liu1987influence, tenailleau1992new, Egerton2002ImprovedSpectra} While some of these issues have been addressed in other studies,\cite{tenailleau1992new, Verbeeck2004ModelSpectra, verbeeck2004model, cueva2012data, Bertoni2008AccuracyQuantification} the established tools are often inadequate to use with noisy spectra containing multiple core-loss edges with limited pre-edge regions, which are common in STEM-EEL spectrum images. In this study, we present a solution to these challenges using a multi-region background fitting method that builds on the approaches presented by Egerton \cite{Egerton2002ImprovedSpectra} and provides simultaneous and robust background subtraction of all separable peaks in the spectrum.

\section{Method}

\begin{figure*}[htb!]
  \centering
  \includegraphics[width = 5in]{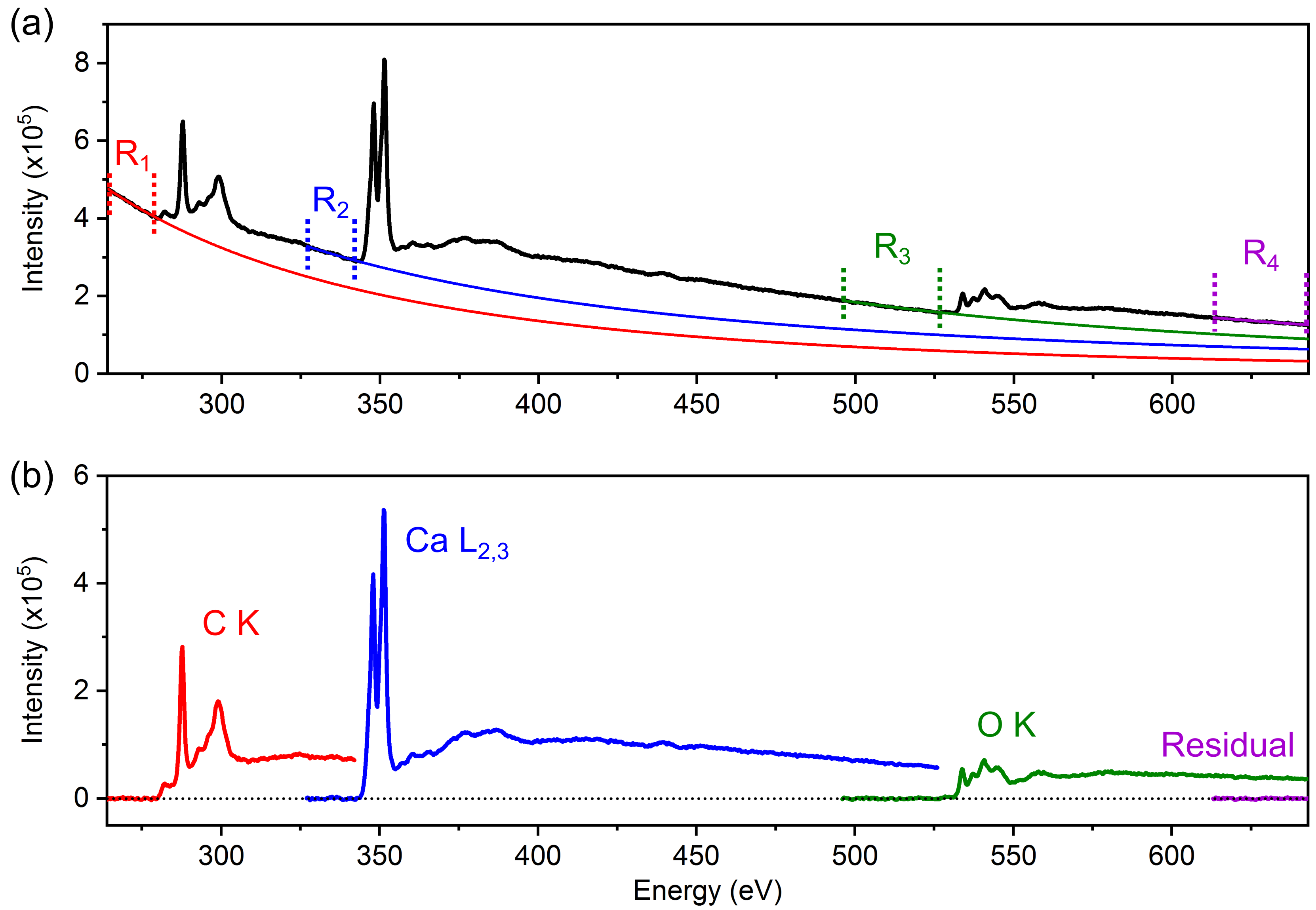}
  \caption{Four-region background subtraction of C $K$, Ca $L_{2,3}$, and O K-edges of calcite in a single EEL spectrum. (a) Raw spectrum (black) showing the background fitting windows ($R_i$ within the dotted vertical lines) and the corresponding extrapolated fits (solid red, blue, green, and purple curves). (b) C $K$, Ca $L_{2,3}$, and O $K$-edges as well as the residual from energy window $R_4$ after background subtraction.}
  \label{multiedge}
\end{figure*}

Just as the pre-edge background under a core-loss EEL edge is well represented by a power law, the post-edge region, beyond any significant extended-loss features, obeys the same functional form.\cite{maher1979functional, Egerton2002ImprovedSpectra} In the method presented here, we use this behavior to constrain and refine the background fitting function for every EEL edge in a spectrum by simultaneously fitting both the pre- and post-edge energy windows with co-dependent functions. Applying this to an isolated EEL edge, the pre-edge background follows a single power law:

\begin{equation}
    f_1(E)=A_1 E^{-r_1} \quad E \in R_1,
    \label{firsteq}
\end{equation}
and the post-edge follows:
\begin{equation}
    f_2(E)=f_1(E) + A_2 E^{-r_2} \quad E \in R_2,
    \label{secondeq}
\end{equation}
where $f_1$ is fit to a 10-30 eV energy window ($R_1$) prior to the edge onset and $f_2$ is fit to a similarly-sized post-edge energy window ($R_2$) beyond any significant edge features.

For an EEL spectrum containing $m$ core-loss EEL edges, a total of $n=m+1$ energy windows must be fit (one pre-edge window for each edge and a final window following the highest-energy edge). The pattern in Eqns. \ref{firsteq} and \ref{secondeq} is extended to accommodate these additional edges by adding a power law term to the fitting function in each subsequent energy window. Accordingly, the function fit to the $j$th energy window is:
\begin{equation}
    f_j(E)=\sum_{i=1}^{j} A_i E^{-r_i} \quad E \in R_j,
    \label{sumeq}
\end{equation}
such that the full piecewise fitting function is:
\begin{equation}
F(E)=
\begin{cases}
f_1(E) & E \in R_1 \\
f_2(E) & E \in R_2 \\
&\vdots\\
f_n(E) & E \in R_n \\
\end{cases}.
\label{piecewise}
\end{equation}

All $A_i$ and $r_i$ are then optimized simultaneously by minimizing the sum-squared error across all windows; this is accomplished with a numerical optimization algorithm such as Nelder-Mead.\cite{nelder1965simplex,press2007numerical} It is worth noting that, due to the large values of $A_i$, it is useful to replace $A_i$ with $a_i=ln(A_i)$.

During fitting, $A_i$ and $r_i$ are subject to constraints:
\begin{equation}
    A_i \geq 0, 
    \label{Aconstraint}
\end{equation}
ensuring the fit function is positive, and
\begin{equation}
    1 \leq r_i \leq 6,
    \label{rconstraint}
\end{equation}
ensuring power-law shape.\cite{maher1979functional, cueva2012data} These constraints prevent non-physical results for the background-subtracted edges such as negative intensities or positive slopes in the post-edge region(s). 

\begin{figure}[!ht]
  \centering
  \includegraphics[width = 3in]{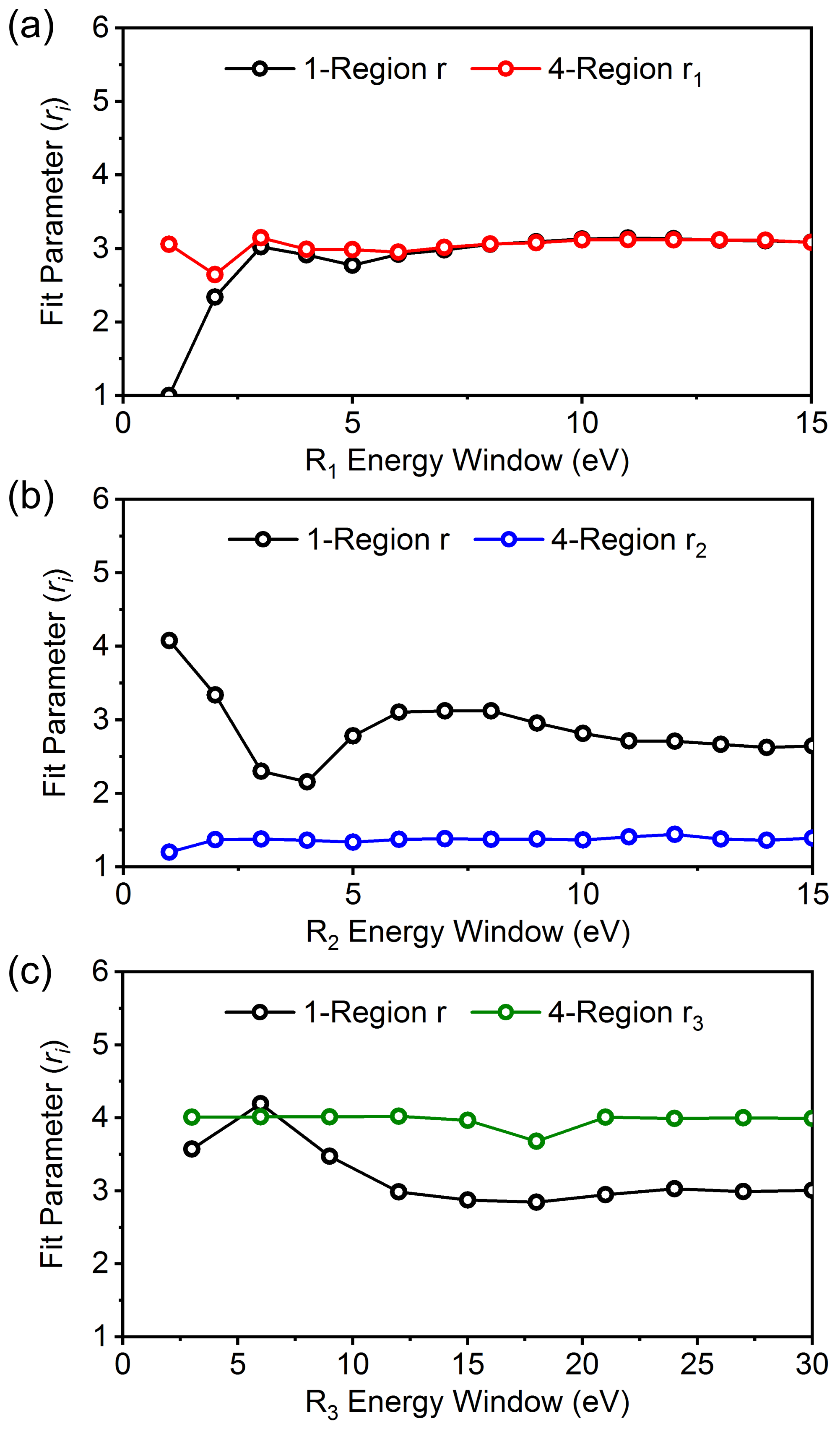}
  \caption{Sensitivity of one- and four-region background fits to the size of fitting window ($R_i$) for the spectrum shown in Figure \ref{multiedge}. (a) Values of fitting parameter $r_1$ for various sizes of energy window $R_1$. The one-region fit is shown in black; $r_1$, the corresponding fitting parameter from the four-region fit is shown in red. (b, c) The procedure from (a) repeated for each of the pre-edge regions of the four-region fit, showing the one-region $r$ and the analogous four-region $r_i$.}
  \label{Windows}
\end{figure}

After fitting, each background function ($f_i$) is extrapolated from the starting point of $R_i$ to the ending point of $R_{i+1}$ and then subtracted from the original spectrum, producing background-subtracted spectra for each EEL edge.

Figure \ref{multiedge} shows a four-region example of the application of this method to an EEL spectrum obtained from a sample of natural calcite (\ce{CaCO3}) prepared by crushing with a mortar and pestle under isopropyl alcohol, and depositing on a holy carbon grid. EEL spectra were acquired from a thin region of the sample over vacuum in an FEI Titan G2 60-300 STEM, operated at 200 kV with a probe semi-convergent angle of 24 mrad and an EEL spectrometer collection angle of 29 mrad. An energy range of 213-724 eV with a 0.25 eV dispersion was used here to capture the C $K$, Ca $L_{2,3}$, and the O $K$ edges in a single spectrum. 

The background fitting windows $R_1=264-279$ eV (pre-C $K$), $R_2=327-342$ eV (pre-Ca $L_{2,3}$), $R_3=496-526$ eV (pre-O $K$), and $R_4=613-643$ eV (post-O $K$) were used, producing a simultaneous background fit for all three edges. For this spectrum, the optimized parameters were: $a_1=30.4, r_1=3.08$; $a_2=19.3, r_2=1.39$; $a_3=36.2, r_3=3.99$; $a_4=32.2, r_4=3.35$.

According to Eqn. \ref{sumeq}, every additional background fitting window adds two more parameters ($r_i$ and $A_i$) that must be optimized. For spectra with many edges, and therefore many background fitting regions, this adds considerable computational time. It is, therefore, important to establish good initial guesses for each parameter. In the case of a well-characterized experimental setup and known sample mass-thickness, reasonable initial guesses for the $r_i^{(0)}$ values may be estimated.\cite{maher1979functional} In situations where experimental conditions cannot be used to inform the initial guesses of $r_i^{(0)}$, an initial coarse fit may be performed with tighter constraints than the final optimization by fitting each window sequentially and \textit{independently}: $A_1^{(0)}$ and $r_1^{(0)}$ are fit over $R_1$ according to Eqn. \ref{firsteq}, then, holding these values constant, $A_2^{(0)}$ and $r_2^{(0)}$ are fit over $R_2$ according to \ref{secondeq}, etc. These coarse approximations of $A_i^{(0)}$ and $r_i^{(0)}$ can then be used as initial guesses in the full, \textit{simultaneous}, optimization of all parameters. While these coarse initial fits are inherently subject to the same pitfalls as one-region power law fits because they are independently optimized for each region, they are good initial guesses and can dramatically reduce the computational time of the final fit.

Because the background fits for each EEL edge are interdependent and simultaneously optimized, a multi-region fit offers increased stability over a one-region fit, making it less sensitive to noise and the size of the fitting windows(s). To demonstrate this, the spectrum shown in Figure \ref{multiedge} was fit with varying fitting window sizes for each $R_i$. Here, the lower energy limit of the fitting window was changed while the upper limit and the rest of the windows were held constant. The fit was re-optimized for each case. This was repeated for a one-region fit over each of the energy windows, and the resultant values of $r_i$ for each case are shown in Figure \ref{Windows}. It is important to note that, aside from the value of $r_1$ in Figure \ref{Windows}a, the absolute values of $r_i$ obtained from the multi-region fit cannot be directly compared with the one-region fit because the fitting function is not the same. However, the consistency of the $r_i$ values directly relates to the stability of the fit.

Even when one fitting window of the multi-region fit is very small ($<10$ eV), such as when EEL edges are close together, the fit remains stable due to the influence of the surrounding windows. This stabilizing effect for the multi-region fit is strongest when the extrapolated backgrounds of the lower energy regions account for the majority of the background under subsequent regions, increasing the interdependence of the fits for each region.

Under certain conditions, it may be advantageous to add a term or otherwise modify Eqn. \ref{sumeq} to account for features beyond the lower range of the spectrum, which can change the shape of the background such that it no-longer obeys a simple power-law. Depending on the specific conditions of the dataset, such modifications can involve altering the first term (i=1) of Eqn. \ref{sumeq} to model the behavior of the bulk plasmon as described by Tenailleau and Martin (1992)\cite{tenailleau1992new}, or including an additional power law term with a fixed r value as described by Cueva et al. (2012)\cite{cueva2012data}. Any modifications made in this way are applied to the fitting functions for all windows in the multi-region fit.

\section{Implementation}

\begin{figure}[!htb]
  \centering
  \includegraphics[width = 3in]{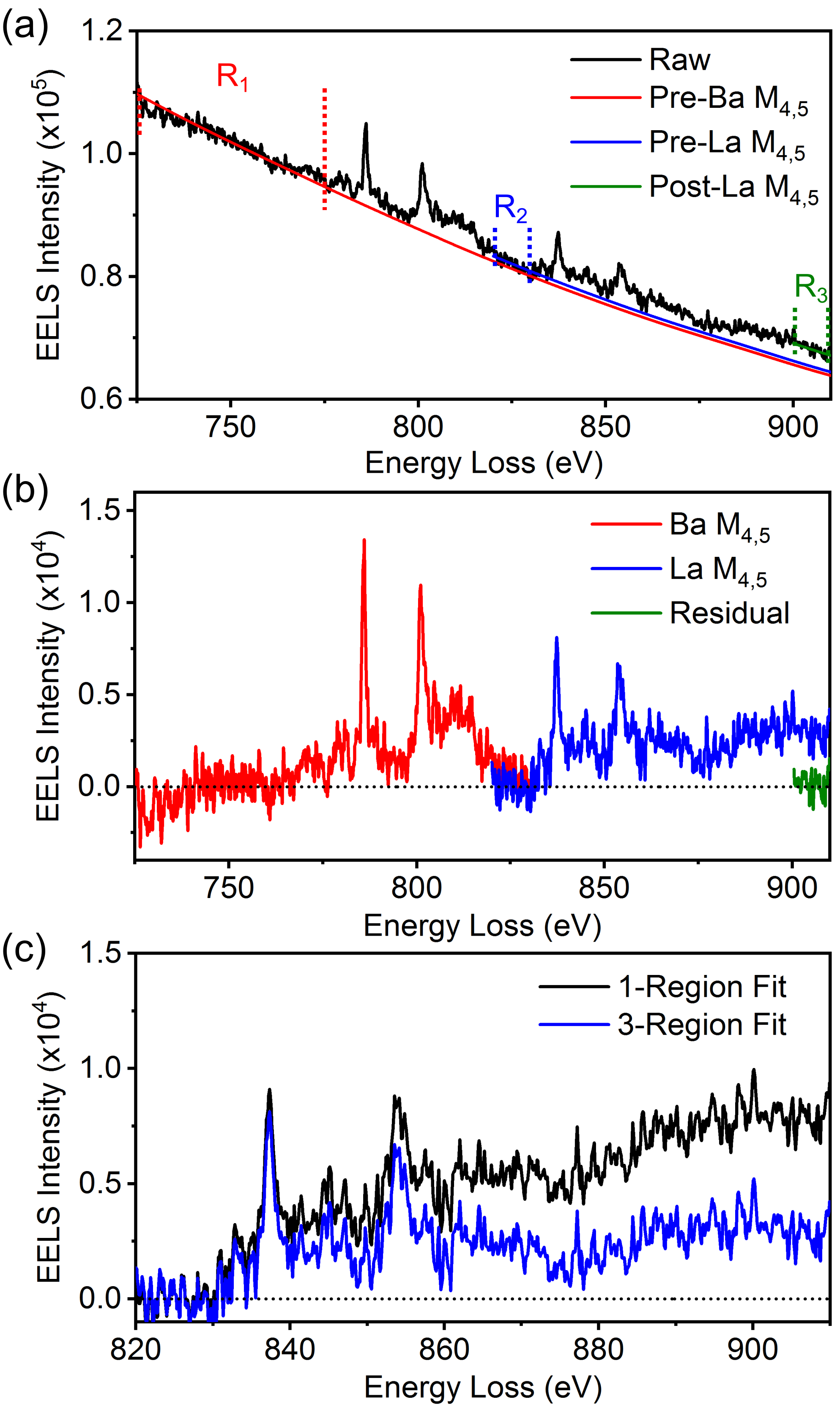}
  \caption{Background subtraction of Ba and La $M_{4,5}$ edges obtained near the interface between BSO and LSSO. (a) Raw spectrum showing the background fitting windows: $R_1=725-775$ eV; $R_2=820-830$ eV; $R_3=900-910$ eV, and the corresponding fits. (b) Background-subtracted Ba and La $M_{4,5}$ edges. (c) Comparison of the La edge produced by this three-region fit with a one-region fit of $R_2$ using the same constraints.}
  \label{BaLasample}
\end{figure}

A multi-region background fit and removal is particularly useful for a STEM-EEL spectrum image or line scan containing multiple core-loss edges. Individual spectra in these datasets are often obtained over very short integration times to mitigate the effects of sample drift and beam damage, resulting in a low signal to noise ratio, which makes conventional background fitting difficult. By making effective use of additional regions of the spectrum, a multi-region fit ensures consistent behavior of every edge in the spectrum. This method facilitates not only better elemental mapping, but also direct comparison of fine structure in each background-subtracted edge.

A linescan across the interface between \ce{BaSnO3} (BSO) and La-doped \ce{SrSnO3} (LSSO) provides a good example of such a situation. In this example, a cross-sectional TEM specimen of a BSO/LSSO heterostructure, grown by hybrid molecular beam epitaxy to study modulation doping at this interface,\cite{prakash2019separating} was prepared using a focused ion beam. A monochromated-EELS linescan was obtained across the interface of the two materials using an FEI Titan G2 60-300 STEM operated at 200 kV with a semi-convergent angle of 17 mrad and an EEL spectrometer collection angle of 29 mrad. A sample spectrum from the interface is shown in Figure \ref{BaLasample}.

The shape and intensity of the La $M_{4,5}$ edges were of particular interest for the analysis of this sample because they were used to determine the concentration and location of La dopants across the interface. In this case, there were multiple factors that had to be considered: the La and Ba $M_{4,5}$ edges exhibited a low signal-to-noise ratio; there is only a very small region ($R_2$) between the Ba and La $M_{4,5}$ edges available for background fitting; and the pre-Ba $M_{4,5}$ background was influenced by the extended-loss features of the Sn $M_{2,3}$ edge. Due to these conditions, a one-region background model was inadequate to produce a reliable fit, necessitating a more robust background model. To overcome these challenges, we used a modified three-region fit with background fitting windows: $R_1=725-775$ eV; $R_2=820-830$ eV; $R_3=900-910$ eV.

A standard Sn extended-loss spectrum from intrinsic \ce{SrSnO3} was background-subtracted and included as an additional term ($A_{Sn} I_{Sn}^{(ref)}$) in the background fitting function. This extra term necessitated a larger window for $R_1$ to capture as much of the Sn character as possible and ensure a stable fit. This modified three-region background fit was used to remove the background from the Ba and La $M_{4,5}$ edges (Figure \ref{BaLasample}). For comparison, a one-region fit for the background under the La $M_{4,5}$ edge using the same fitting window ($R_2$) yielded a considerably different and non-physical result (the extended-loss intensity grew rather than obeying a power-law), as shown in Figure \ref{BaLasample}c.

\begin{figure*}[!ht]
  \centering
  \includegraphics[width = 5in]{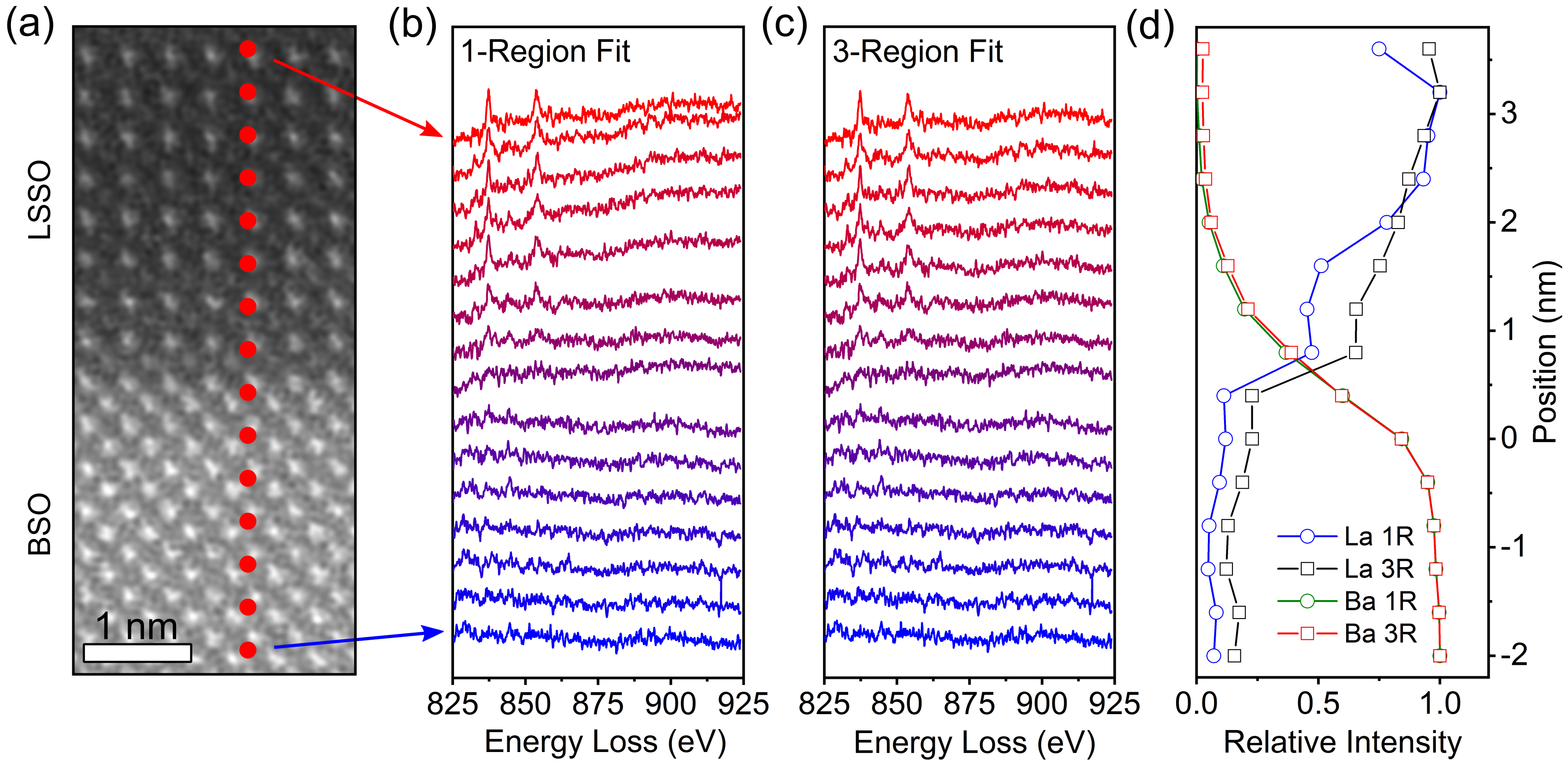}
  \caption{STEM-EELS line scan across the interface between BSO and LSSO. (a) HAADF-STEM image of the region of interest with red dots indicating the approximate locations of each data point in the line scan. (b) La $M_{4,5}$ edges for each point on the line scan using one-region background subtraction. (c) La $M_{4,5}$ edges for each point on the line scan using the three-region background subtraction shown in Figure \ref{BaLasample}. (d) Normalized relative intensities of Ba $M_{4,5}$ and La $M_{4,5}$ edges obtained by scaling standard spectra obtained from the same sample to the background-subtracted edges using one- and three-region background subtraction.}
  \label{FullLaFit}
\end{figure*}

When extended to the rest of the spectra in the line scan (Figure \ref{FullLaFit}), the aberrant behavior of the one-region background subtraction is even more apparent. While the three-region fit produced a very consistent and sensible shape in the La $M_{4,5}$ edge, the behavior of the one-region fit varied significantly along the line scan. Specifically, the one-region fit over-subtracted the background at the BSO end of the line scan (causing the background-subtracted spectra to be negative) and under-subtracted the background at the LSSO end of the line scan (as discussed for the example in Figure \ref{BaLasample}c). The degree of this over- and under- subtraction caused the shape and quantification of the La $M_{4,5}$ edge to vary dramatically between spectra.

To quantify the relative content of La and Ba across the interface while minimizing the influence of noise, a standard La $M_{4,5}$ spectrum was obtained from the LSSO and background-subtracted using the same conditions as the rest of the line scan. This standard was then scaled to the background-subtracted La $M_{4,5}$ edge at each point using linear least squares fitting. This procedure was repeated with the Ba edge to produce the concentration profiles shown in Figure \ref{FullLaFit}d. Although the one- and three-region cases agree well on the Ba profile, the comparative consistency of the three-region background subtraction resulted in a much smoother curve for the La profile.

It is worth noting that for the low-concentration points along the line scan, the three-parameter background subtraction case consistently over-estimated the concentration of each element, though this was more noticeable for La. This is because the third region ($R_3$) was forced to follow power law behavior to minimize fitting error; that is, it could not be negative, and it had to maintain a negative slope in that region. Lacking such constraints, the one-region background subtracted spectra for the same points often became negative in the $R_3$ region. In this case, minor extended loss features from the Ba edge may have exacerbated the issue, but this tendency for a multi-region background fit to under-subtract the background should be taken into consideration when working with low signal-to-noise data.

\section{Conclusion}

The multi-region background fitting method presented in this study offers more computationally expensive, but more reliable background subtraction than customary one-region power-law fits. By utilizing both the pre- and post-edge regions of each edge in the spectrum to determine the best-fit background, it ensures consistent behavior of the background-subtracted spectra. This method is particularly well suited for simultaneous fitting of multiple edges because it not only benefits from the additional background regions, it also inherently yields background-subtracted spectra for each edge.

\section{Acknowledgments}
This work is supported in part by SMART, one of seven centers of nCORE, a Semiconductor Research Corporation program, sponsored by National Institute of Standards and Technology (NIST), and by UMN MRSEC program under award no. DMR-1420013. This work utilized the College of Science and Engineering (CSE) Characterization Facility, University of Minnesota (UMN), supported in part by the NSF through the UMN MRSEC program. H. Y. acknowledges a fellowship from the Samsung Scholarship Foundation, Republic of Korea. J. T. H. acknowledges support from a Doctoral Dissertation Fellowship received from the graduate school at the University of Minnesota.

\newpage

\bibliography{references}

\end{document}